# Mesonic Super Čerenkov-like effects in hadronic and nuclear media

D. B. Ion[1,2) and M.L. Ion[3)

[1) National Institute for Physics and Nuclear Engineering Horia Hulubei, IFIN-HH,
Bucharest, P.O.Box MG-6, Magurele Romania

[2) Academy of Romanian Scientist (A.O.S.R.)

[3) Faculty of Physics, Bucharest University, Bucharest, Romania

**Abstract**: Generalized mesonic Super-Čerenkov Radiations (SČR) are investigated. The energy behavior of the pionic refractive index in the low energy pionic SČR-sector is presented. We estimated that the true coherent SČR-pion emission is possible mainly in the SČRS-energy bands 190-315 MeV for all pions; 910-960 GeV only for positive pions, and 80-1000 GeV for all pions, in certain nuclear reactions. We predicted that SČR-pionic band will be enlarged for the pion energies higher than 80 GeV. The strong correlations between angle of SČR-pion emission and (meson and projectile)-energies are evidentiated. The spectral distributions of the SČR-pions are presented and the position of their maxima are estimated. The agreement with the available experimental data is discussed.

We recall here that idea that meson production in nuclear interactions may be described as a process similar to the Čerenkov radiation has been considered by Wada (1949), Ivanenko (1949) Blohintev and Indenbom (1950), Čzyz, Ericson, Glashow (1959), Smrz (1962) and D. B. Ion (1969-1970). For the many detailed results on mesonic Čerenkov-like effect see Refs. [1-3,5,9-13] presented in Fig.1 while for the generalized Super-Cerenkov see Refs. [14-15]. In Ref. [1] from Fig.1 D.B. Ion developed a general classical and quantum theory of the mesonic Čerenkov-like radiation in hadronic and nuclear media. Moreover, the vector-mesonic Čerenkov-like radiation as well as baryonic Cerenkov-like effects in nuclear and hadronic media were also introduced for the first time in Ref. [1-2,Fig.1]. Then, there were predicted completely the properties of the mesonic Čerenkov-like radiation in the case when the mesonic refractive index is given by a single pole approximation. Then, they obtained a good agreement with the integrated cross section of the single meson production in the hadronic collisions (see Fig.9 from Ref. [1]).

In 1990-1995, we have extended [1-7] these ideas to the nuclear media where the pionic (NPIČR) and gamma Čerenkov radiation (NGČR) should be possible to be emitted from charged particles moving through nuclei with a velocity larger than the phase velocity of photons or/and pions in the nuclear media. The refractive indices of the gamma ($n_\gamma$), meson ($n_\pi$), nucleon ($n_N$), was calculated by using Foldy-Lax formula [8] and the experimental pion-nucleon cross sections combined with the dispersion relations predictions, the refractive index of pions in the nuclear media have been calculated (see Fig.2a,b). Then, the detailed predictions for the spontaneous pion emission as *nuclear pionic Čerenkov radiation* (NPIČR) inside the nuclear medium are obtained and published in Refs. [3-4]. These main predictions and conclusions obtained in this way for the low-energy SČR-pionic sector can be summarized as follow (see also Fig.4):

(i) The energy behavior of the pionic refractive index is presented in Fig. 2a;

(ii) The true coherent pion emission as *nuclear pionic Čerenkov-like radiation* (NPIČR) is possible in the following three energy bands (see Fig. 2b):

ČB1-NPIČR band for: $190\,MeV \leq \omega \leq 315\,MeV$ for all $\pi^{\pm,0}$

ČB2-NPIČR band for: $910\,MeV \leq \omega \leq 960\,MeV$ only for $\pi^+$, and

ČB3-NPIČR band for: $80\,GeV \leq \omega \leq 1000\,GeV$ for all $\pi^{\pm,0}$

in the nuclear reactions such as: $N + {}^{208}Pb \Rightarrow \pi N + {}^{208}Pb$ .

Here, it is important to note that, for the nucleon laboratory momenta $p_{LAB} \geq 80\,GeV/c$ , we predict that the all above SČR-pionic bands will be enlarged since the physical domain is given by

$$v_{\pi\,ph}(\omega)/v_1\,\mathrm{Re}\,n_1(E_1) \leq \cos\theta_{N\pi} \leq 1$$

and $\mathrm{Re}\,n_N(E_N) \geq 1$ (see the results from Fig. 4).

(iii) The NPIČR-pions must be coplanar with the incoming and outgoing projectile possessing a strong correlation between the angle of emission $(\theta, \omega_m)$ and the pion ($\omega$) and projectile ($T_p$) energies (see Fig.3).

(iv) For the ČB1 the NPIČR-differential cross section are peaked (see Ref. [4]) at the energy $\omega_m = 260$ MeV for ČB1 band and $\omega_m = 930$ MeV for ČB2 band when the absorption is neglected and the peak position is shifted up to $\omega_m = 240$ MeV for ČB1-band when the absorption is taken into account. As we already mentioned in Ref. [1], these predictions were experimentally confirmed (see Fig. 10 in [1]) by the Dubna group (see E.K. Sarkisyan et al. Phys. Lett. B471 (1999) 257). So, they obtained a good agreement with the position and width of the first pionic Cerenkov-like band predicted in Ref. [4].

**Fig.1**: Selected bibliography for introduction in the generalized Super-Čerenkov-like radiations.

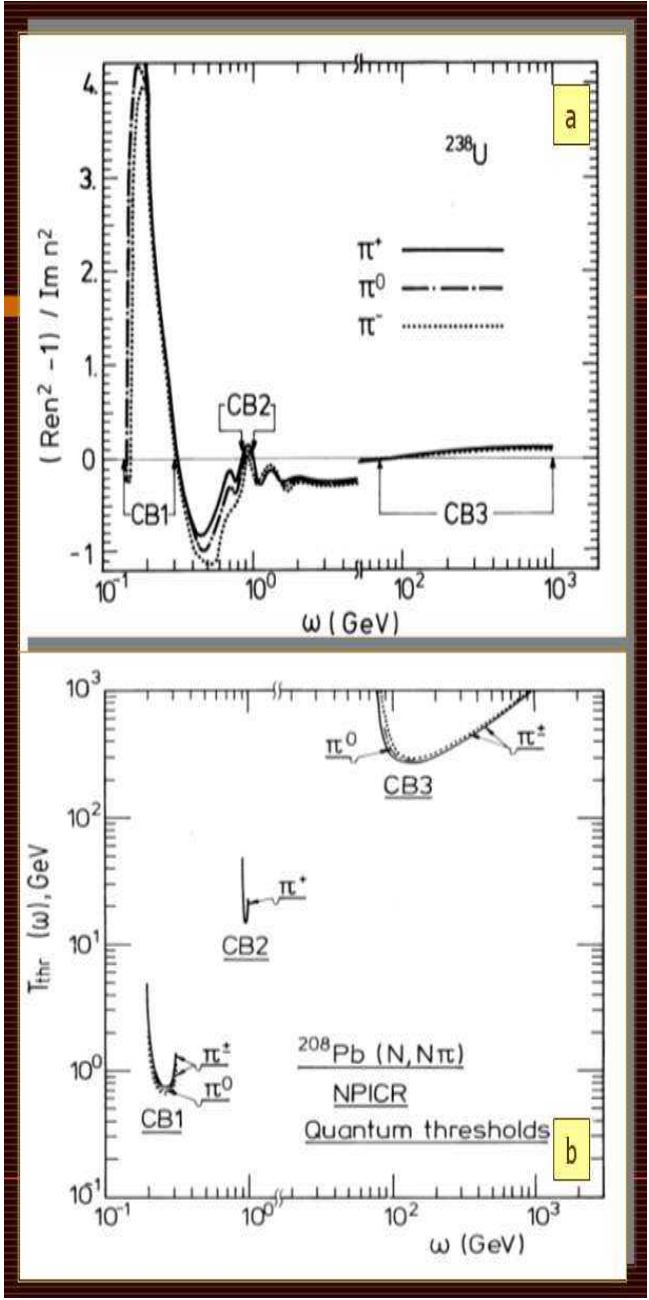

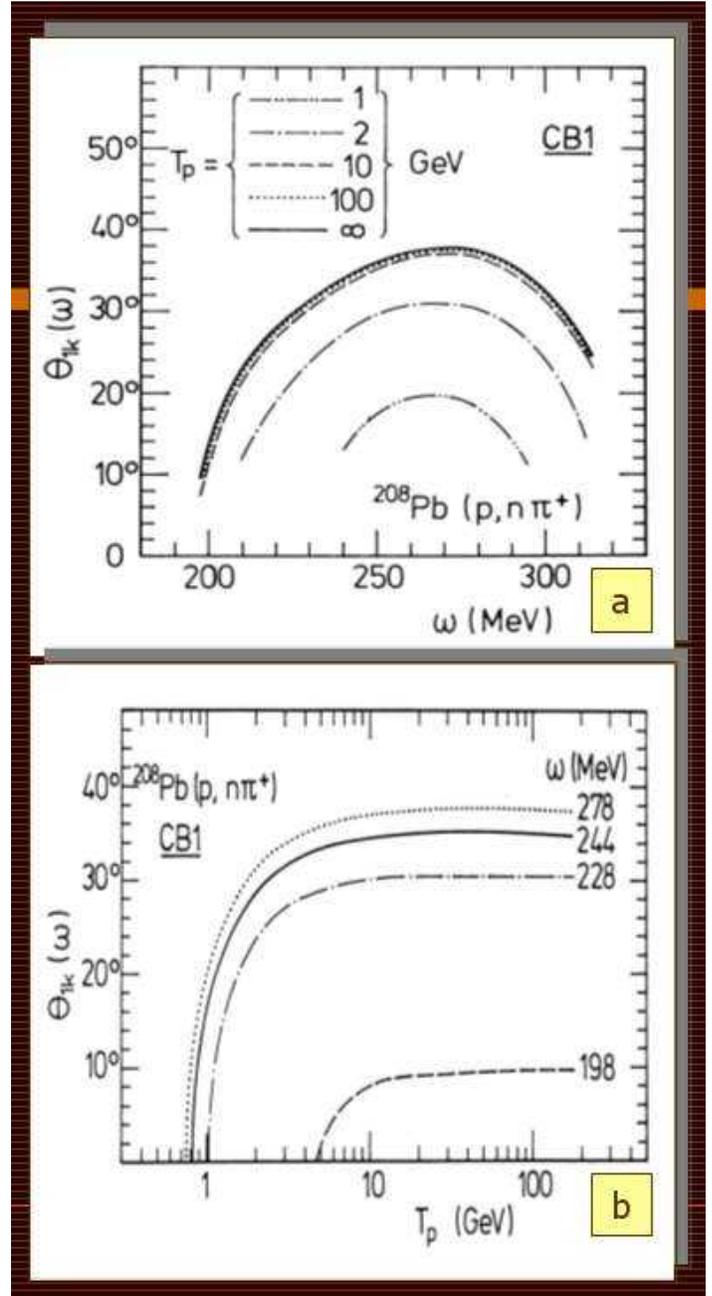

**Fig. 2**: (a) Pionic refractive index, (b) The SČR-energy thresholds.

**Fig. 3**: $[(\theta_{1k}, \omega)$ and $(\theta_{1k}, T_p)]-$ angle-energy correlations.

(v) A summary of the theoretical results on pionic Super-Čerenkov-like radiation in (hadronic or nuclear) media is presented in Fig.4. The factor S is the spin factor while $\Theta(1-\cos\theta_{SC})$ is the Heaviside step function. In fact the entire quantum theory of the exotic decay: $B_1(\vec{p}_1, E_1) \rightarrow \pi(\vec{k}, \omega) B_2(\vec{p}_2, E_2)$, where $B_1$ and $B_2$ are spin ½ baryons, can be developed just as in Ref. [4]. So, by using the inequality: $\cos\theta_{SC} \approx v_{\pi\,ph} v_{Bph} \leq 1$, two general SČR-coherence conditions corresponding to the (mesonic and baryonic)- Čerenkov-like effects, are found (see Fig.4) as two natural extremes of the same spontaneous particles decays in medium.

(vi) Next, it is important to remark that the baryonic SČR-sector will appear especially for incident nucleons with $p_{LAB}$ higher than 100 GeV/c where we found that the baryonic SČR-conditions: and $R\, v_{N\,ph}(E_2)/v_1\,\mathrm{Re}\,n_1(E_1) \leq \cos\theta_{12} \leq 1$ and $\mathrm{Re}\,n_N(E_N) \geq 1$, are satisfied with high accuracy.

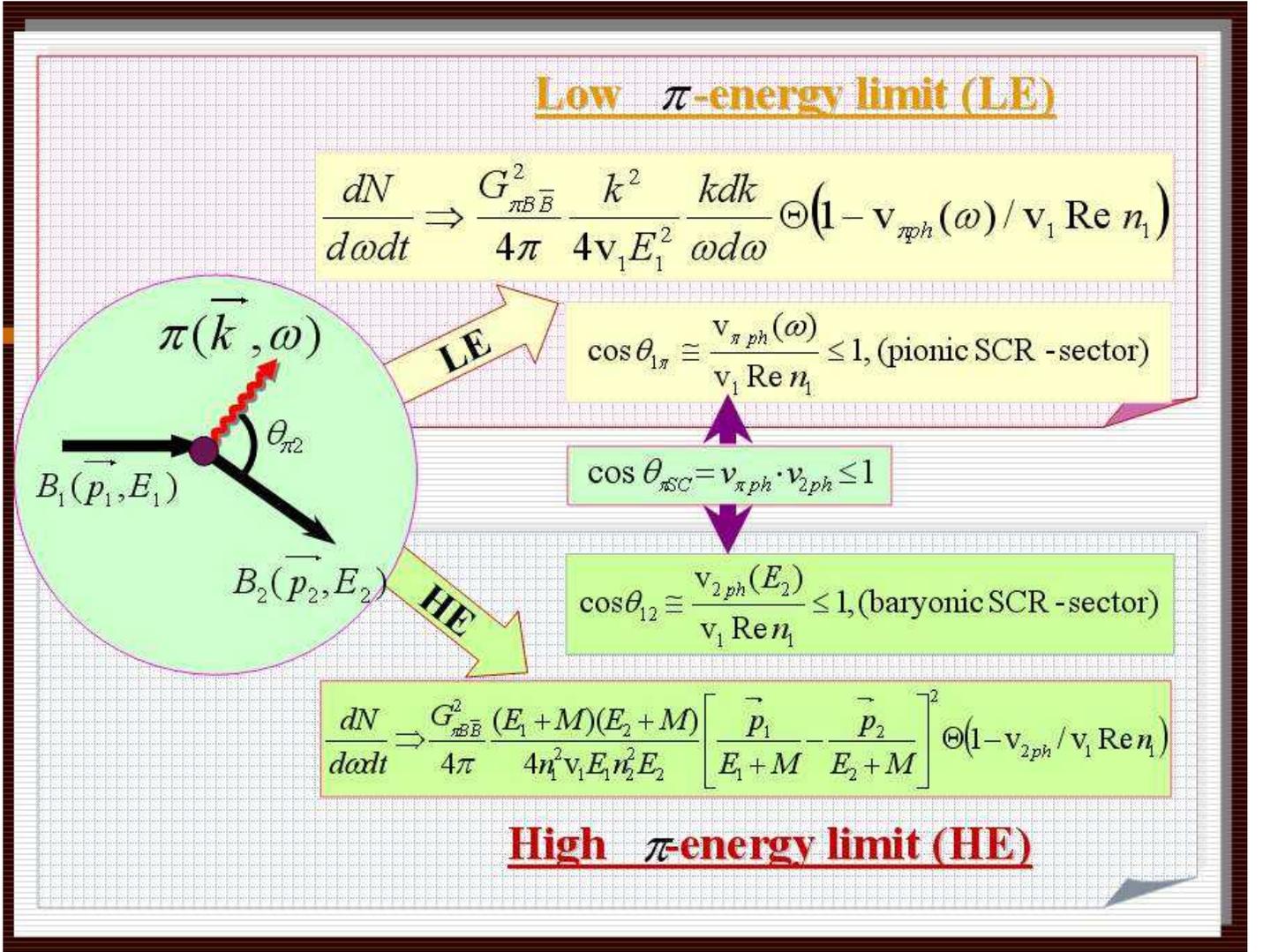

**Fig. 4:** Summary of the theoretical results on pionic Super-Čerenkov-like radiation in (hadronic, nuclear) media.

(vii) It is well known that the recent RHIC experiments [6, 7] have shown two bump structure of the azimuthal distributions near the away-side jets. This structure was interpreted by Dremin [8] as being the signature of the Čerenkov gluons. But, it is easy to see that, these two bump distributions can be interpreted in more exact way: as signature of the two components of the SČR-gluons. However, the more realistic interpretation of these experimental results as signature of the generalized mesonic SČR-effects cannot be avoided. Of course more theoretical and experimental investigations are necessary to clarify the problems of the generalized SČR-gluons emissions in hadronic media.

**Acknowledgments**

This research was supported by CNCSIS under contract ID-52-283/2007.